\documentclass[12pt,twoside]{article}
\usepackage{fleqn,espcrc1}

\usepackage{graphicx}
\usepackage[figuresright]{rotating}

\newcommand{\AmS}{{\protect\the\textfont2
  A\kern-.1667em\lower.5ex\hbox{M}\kern-.125emS}}

\title{Solving the Schwinger-Dyson equation for a scalar propagator in Minkowski space}

\author{V.~\v{S}auli\address{Faculty of Mathematics and Physics, Charles University,
          V Hole\v{s}ovi\v{c}k\'ach 2, \\ CZ 180 00, Prague, Czech Republic}
          \thanks{PhD student at Nuclear Physics Institute in  \v{R}e\v{z} near Prague}
        and
        J.~Adam\address{Nuclear Physics Institute, Academy of Sciences of the Czech
        Republic, \\
         \v{R}e\v{z} near Prague, CZ-25068, Czech Republic}
          \thanks{This work is supported by the grant GA CR 202/00/1669}  }

\begin{document}

\maketitle

\begin{abstract}
The Schwinger-Dyson equation for a scalar propagator is solved in Minkowski
space with the help of an integral spectral representation, both for spacelike and
timelike momenta. The equation is re-written into a form suitable for
numerical solution by iterations. This procedure is described for a
simple unphysical Lagrangian with a cubic interaction, with future extensions
to more realistic theories in mind.
\end{abstract}

\section{INTRODUCTION}

The infinite tower of integral Schwinger-Dyson equations (SDEs) links n-point
Green's functions ($n = 2, 3 \dots$) of a quantum-field theory. Their exact
solution would provide complete information on the physics of the theory, including
its non-perturbative regime. In practice, the system of SDEs has to be truncated
and closed by making assumptions about the driving term, which contains more complicated
Green's functions determined by equations that were thrown away.
Most often only the simplest equation for the 2-point Green's function(s)
(one particle propagator) is retained, its kernel contains n-point ($n= 3,\dots$)
vertex the full form of which is unknown. Hence some physical ansatz has to be used and
then the equations can be solved (usually numerically). One hopes that the solution
provides some useful information on the behavior of the theory, in particular in
the non-perturbative region. It is certainly interesting to compare the solution
with results of alternative non-perturbative techniques \cite{Cetin}.

In most papers on the solution of the SDE, the Wick rotation from Minkowski to Euclidean
space is employed in order to escape singularities of the kernel inherent to physical
Green functions. We instead attempt to find the solution directly in Minkowski space by
making use of the spectral representation of the Green's functions based on their
known (or assumed) analytical properties. In this contribution we will for simplicity
illustrate this method of solution with a simple example of self-interacting scalar fields
with super-renormalizable unphysical $g \Phi^3(x)$ coupling. Possibilities of extending
the technique to more realistic theories are briefly discussed.

\section{FORMALISM}

Our approach is a straightforward extension of the spectral decomposition method of
Kusaka {\em et al.} \cite{Kusaka}, developed for the solution of the Bethe-Salpeter equation for
scalar bound states. The Green's functions -- one scalar particle propagator in our
case -- are written as spectral integrals over some weight functions. Then we put these
parameterizations and an expression for the vertex function into the SDE, combine
denominators with the help of the usual Feynman parameterization, integrate over the loop
momentum/momenta and obtain the real integral equations for the weight functions depending
on spectral variables. These equations are free of singularities and can be solved by iterations.

The generic spectral decomposition of the dressed renormalized
scalar propagator reads:
\begin{equation}
 G(p^2) = \int d \alpha \frac{\sigma(\alpha)}{p^2- \alpha- i \epsilon} \  , \ \ \
 \sigma(\alpha)= \delta(\alpha-m^2)+ \rho(\alpha) \ ,
\label{genspec}
\end{equation}
where $\sigma(\alpha)$ is a positive spectral function. Here, we
assume that $\sigma$ has a singular contribution due to the
propagation of physical particle with the mass $m$ and a regular
positive smooth part $\rho(\alpha)$ which starts at the
two-particle threshold. This assumption means that the particle
spectrum of the system is essentially perturbative, e.g., there is
no confinement and also the contribution of possible bound states
below $\alpha_{th}= 2m$ is neglected (the latter appears as
corrections to the vertex from the poles in the 4-point Green's
function).

The form of the SDE for an one particle propagator depends on the form of the interaction. For our
toy model with ${\cal L}_{int}= - g \Phi^3(x)$ and introducing the renormalized
(by on-mass shell subtraction) self-energy $\Pi_R(p^2)$ we get
\begin{eqnarray}
G^{-1}(p^2)&=& Z(p^2-m_0^2)- \Pi(p^2)= p^2- m^2- \Pi_R(p^2)\, ; \
\Pi_R(m^2)= \frac{d}{d p^2} \Pi_R(m^2)= 0 \ , \label{self}\\
\Pi(p^2) &=& i\, \frac{S g^2}{(2\pi)^4}\,  \int d^4q\, \Gamma(p,q)\,G((p-q)^{2})\,G(q^{2}) \, ,
\label{sde}
\end{eqnarray}
where $m_0$ is a bare mass, the residue of $G(p^2)$ at the pole
$p^2=m^2$ equals 1; $S= 18$ is a combinatorial factor, $Z$ is the renormalization function of the
$\Phi(x)$ field, which is finite for the cubic interaction. The adopted renormalization procedure
suggests the following spectral decomposition for $\Pi_R(p^2)$:
\begin{equation}
\Pi_R(p^2) = (p^2-m^2)^2\, \int d \alpha\, \frac{\rho_{\pi}(\alpha)}{p^2- \alpha- i \epsilon} \  .
\label{selfdis}
\end{equation}
We will mostly work in the bare vertex approximation
 $\Gamma(p,q) =1$.  From the imaginary part of $G= G_0 + G_0\Pi_R G$ and from the
dispersion relations of Eqs.\ (\ref{genspec}) and (\ref{selfdis}) follows:
\begin{equation}
\rho(\omega) = \rho_{\pi}(\omega)+ (m^2-\omega)\, P \int_{4m^2}^{\infty} d \alpha \,
\frac{\rho_{\pi}(\omega)\rho(\alpha)+\rho_{\pi}(\alpha)\rho(\omega)}{\alpha-\omega} \, .
\label{sderho}
\end{equation}
From the SDE (\ref{sde}) we get the second relation between $\rho_\pi$ and $\rho$:
\begin{eqnarray}
 \rho_{\pi}(\omega)&=& \frac{S g^2}{(2\pi)^4}\, \frac{1}{(\omega-m^2)^2} \left[ \sqrt{1-4m^2/\omega}
 +   \int d\alpha\, T(\alpha,m^2,\omega) \rho(\alpha)  \right. \nonumber\\
 && \hspace*{5truecm}  \left. + \int \int d\alpha_1 d\alpha_2\, T(\alpha_1,\alpha_2,\omega)
 \rho(\alpha_1)\rho(\alpha_2)  \right] \, ,
\label{rhopi}
\end{eqnarray}
where $T(\alpha_1,\alpha_2,\omega)$ is a purely kinematical
function which also determines the integration bounds. The set of
equations (\ref{sderho},\ref{rhopi}) is solved by iterations.

The method works in essentially the same way also for more complicated theories. Of course, if
there is more than one type of particles, one gets a set of coupled integral equations.
In more complicated theories it might not be always possible to integrate out all Feynman
parameters. Then kinematical functions analogous to $T(\alpha_1,\alpha_2,\omega)$ in Eq.\ (\ref{rhopi})
are expressed in terms of the remaining integrations. That might slow down and complicate
the solution, in particular one has to determine numerically the bounds for integrals over the
spectral parameters $\alpha$.

\section{RESULTS}

The iterations work very well for small $\lambda= S g^2/((2\pi)^4\, m^2)$, but in that region
the propagator  is not far from the free one. With increasing $\lambda$ the rate of convergence
is slowing down rather fast, then for a certain value $\lambda_{crit} \sim 2.5$ it breaks down. We do not
know at this stage the meaning of this critical value of the coupling constant, neither for our
unphysical example of the scalar cubic interaction nor for other theories (e.g.\ $\Phi^4$).
We even do not know whether the SDE indeed does not have a (formal) solution for larger
$\lambda$ or whether it is just a failure of our numerical method. We would only observe that a
very similar behavior was found in \cite{Ahlig}, where a similar coupling $\Phi^2 \Psi$ was
considered. In our case it is also possible (in a slightly modified formulation) to calculate
the field renormalization function $Z$ which is finite. It appears that very close to
$\lambda_{crit}$ (within 5 per cent) this function goes through zero (another sign of
pathological behavior). We are aware of the fact that the $\Phi^3$ model is in
fact not defined at all \cite{Baym}, hence its study should be viewed as purely methodological
and one has to be very cautious in making any generalization. It might be nevertheless interesting
that when one includes one loop corrections to the vertex function,  $\lambda_{crit}$ drops
by a factor of about 2.

Figure 1 shows the weight functions $\rho(\alpha)$ from below for
$\lambda= 0.1, 0.5, 1.0, 2.2, 2.5$. The last line already shows wiggles,
precursors of the numerical breakdown. Figure 2 displays the self-energies below
threshold $s=4m^2$ from above for $\lambda= 0.25, 1.0, 1.5, 2.2, 10.0$,
compared with those obtained by Dyson (bubble) summations (for the last value $\lambda=10.0$
we do not get a solution of the DSE). For small
$\lambda$'s the Dyson summation approximates our full solutions rather well, for $\lambda$
close to $\lambda_{crit}$ they deviate by up to 25 per cent.

\section{CONCLUSION}

We have developed and tested in a cubic scalar toy model the method of
solving DSE's in Minkowski space. Several applications of the method are in progress:
studies of the scalar $\Phi^4$ theory, of 3+1 QED in a quenched rainbow approximation, and of a
theory with $\bar{\psi} \gamma_5 \psi \Phi$ fermion-pseudoscalar coupling. The solution of SDEs
in the described framework seems to be feasible for all of these models, even with more complicated
structure of equations employed. When one tries to include more loops, much more
painful algebraic manipulations has to be done to bring equations for weight functions
into manageable form. We hope to prove by this further tests that this method is competitive,
compared to solutions in Euclidean space.

\begin{center}
\includegraphics[height=10.5cm,angle=270]{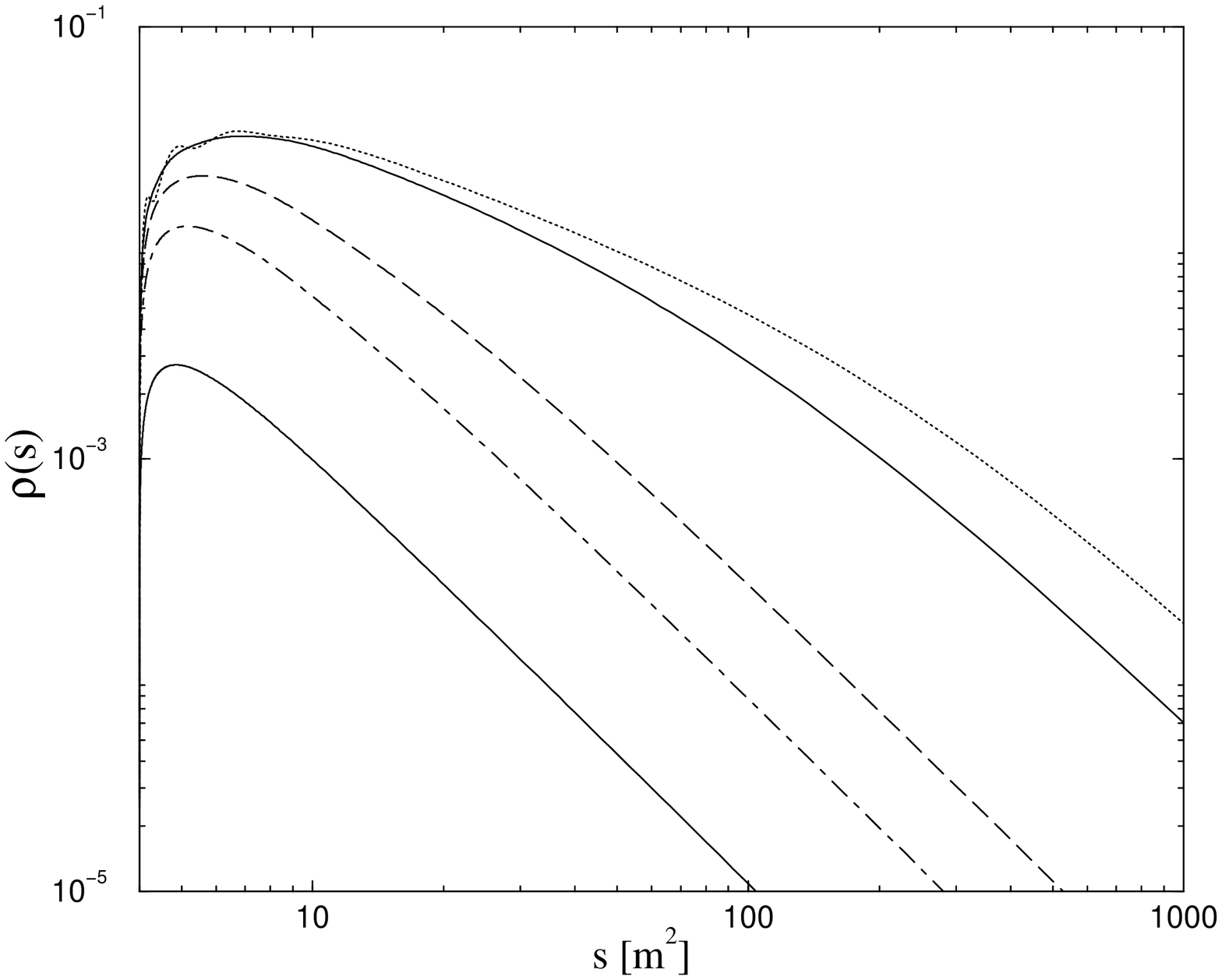} \\
Fig.1
\end{center}

\begin{center}
\includegraphics[height=10.5cm,angle=270]{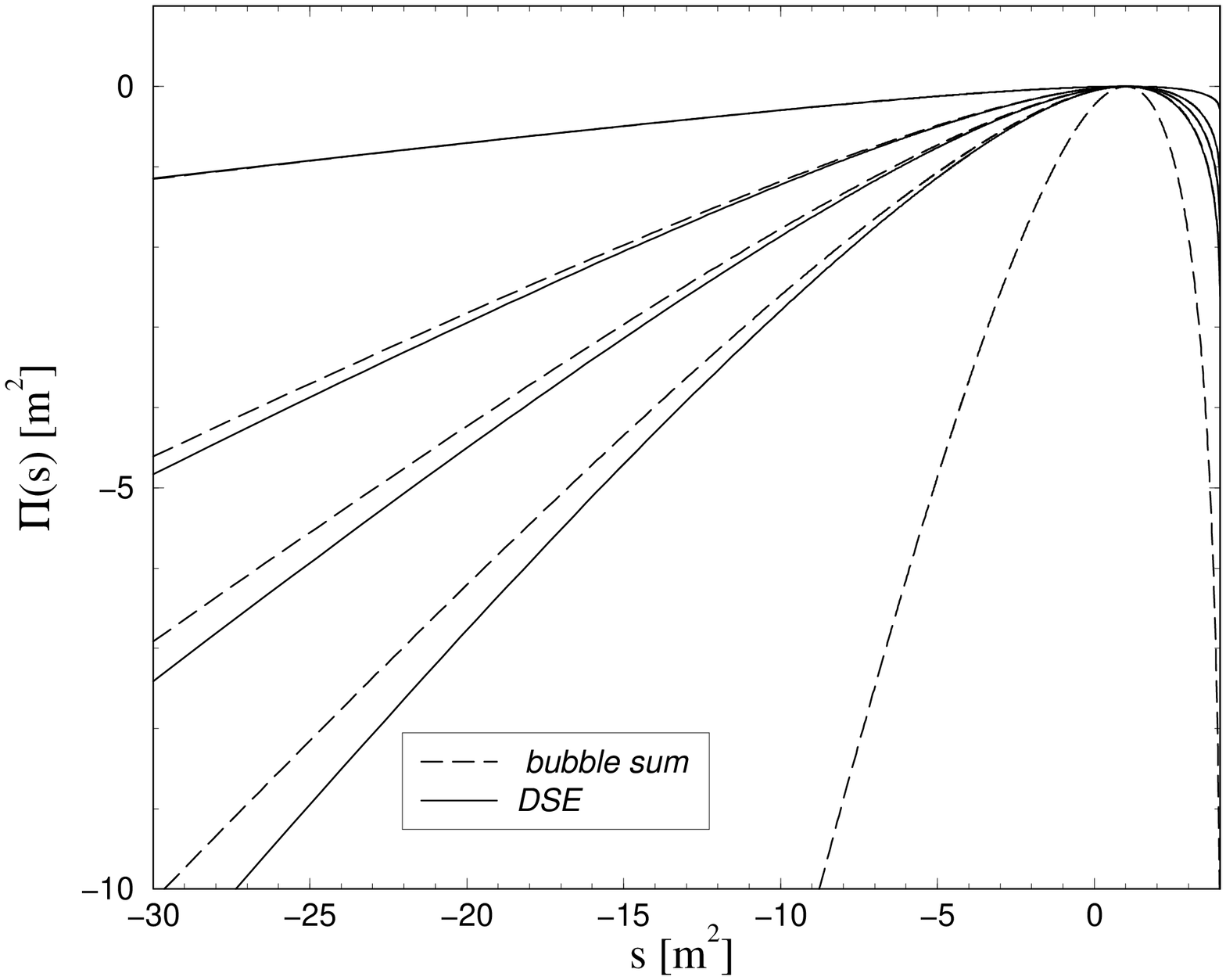} \\
Fig. 2\  Curves in figures are described in the text.
\end{center}

\end{document}